\documentclass[%
 reprint,
 amsmath,amssymb,
 aps,
prl,twocolumn
]{revtex4-1}
\usepackage{graphicx}
\usepackage{bm}
\usepackage{float}
\usepackage{tabularx}
\def \gese2{Ge(Se$_{1/2}$)$_4$}
\def \ge2se3{Ge$_2$(Se$_{1/2}$)$_4$} 
\def \dE{$\delta$E}
\def \w0{$\sigma(0)$}
\begin{document}


\title{Amorphous to amorphous insulator-metal transition in GeSe$_3$:Ag glasses}

 \author{Kiran Prasai}
 \author{Gang Chen}
 \author{David Drabold}%
 \email{drabold@ohio.edu}
\affiliation{%
 Ohio University, Department of Physics and Astronomy, Athens, Ohio, 45701
}%

\date{\today}

\begin{abstract}
We study an insulator-metal transition in a ternary chalcogenide glass (GeSe$_3$)$_{1-x}$Ag$_x$ for $x$=0.15 and 0.25. 
The conducting phase of the glass is obtained by using ``Gap Sculpting" (Prasai et al, Sci. Rep. 5:15522 (2015)) and it is observed that
the metallic and insulating phases have nearly identical DFT energies but have a conductivity contrast of $\sim 10^8$.
The transition from insulator to metal involves growth of an Ag-rich phase accompanied by a depletion of tetrahedrally bonded \gese2 in the host network.
The relative fraction of the amorphous Ag$_2$Se phase and GeSe$_2$ phase is shown to be a critical determinant of DC conductivity.
\end{abstract}

\pacs{Valid PACS appear here}
\maketitle

Metal-Insulator transitions (MIT) and their associated science are among the cornerstones of condensed matter physics \cite{mott2012}.
In this Letter, we describe the atomistics of a technically important but poorly understood MIT in GeSe:Ag glasses, a prime workhorse of conducting bridge memory (CBRAM) devices \cite{patent1,valov2011}.
By {\it design}, we construct a stable conducting model from a slightly favored insulating phase.
Predictions are made for structural, electronic and transport properties. We demonstrate the utility of our ``Gap sculpting" method \cite{prasai2015} as a tool of Materials Design.

We report metallic phases of amorphous (GeSe$_3$)$_{1-x}$Ag$_{x}$ at $x=0.15$ and $0.25$. 
These are canonical examples of Ag-doped chalcogenide glasses, which are studied in relation to their photo-response
and diverse opto-electronic applications \cite{kolobov2006,inbook2}. 
Ag is remarkably mobile making the material a solid electrolyte and is known to act as ``network-modifier" in these glasses and alter the connectivity of network.
Experiments have shown Se rich ternaries ((Ge$_y$Se$_{1-y}$)$_{1-x}$Ag$_x$ with y $< 1/3$) to be phase-separated into
Ag-rich Ag$_2$Se phase and residual Ge$_t$Se$_{1-t}$ phase \cite{mitkova1999}.

Using first-principles calculations, we show that stable amorphous phases with at least $\sim 10^8$ times higher electronic conductivity
exist with only small ($\approx 0.04$ eV/atom) difference in total energy. 
These conducting states present the same basic structural order in the glass, 
but have a higher relative fraction of an {\it a-}Ag$_2$Se phase compared to the insulating states.
It is known that amorphous materials are characterized by large numbers of degenerate conformations that are mutually
accessible to each other at small energy cost, but those usually have identical macroscopic properties. 
The remarkable utility of these materials accrues from states with distinct properties, nevertheless readily accessible to each other.

We discover the conducting phase of GeSe$_3$Ag glass by {\it designing} atomistic models with a large density of states (DOS) near the Fermi energy \cite{prasai2015}. 
This is achieved by utilizing Hellmann-Feynman forces from the band edge states. 
These forces are used to bias the true forces in {\it ab initio} molecular dynamics (AIMD) simulations
to form structures with a large DOS at the Fermi level. The biased force on atom $\alpha$, $F^{bias}_{\alpha}$, is obtained by suitably summing
Hellmann Feynman forces for the band edge states (second term in Eq. \ref{eq_a}) with the total force from AIMD calculations, $F^{AIMD}_{\alpha}$.
\begin{equation}\label{eq_a}
{F}^{bias}_{\alpha} =  {F}^{AIMD}_{\alpha}+\sum \limits_{i} \gamma_{i}  \langle \psi_{i}| \frac{\partial H}{\partial R_{\alpha}}|\psi_{i} \rangle 
\end{equation}
\noindent  
Here,  $\gamma$'s set the sign and magnitude of the HF forces from individual states {\it i}. To maximize the density of states near $\epsilon_F$,
gap states closer to the valence edge will have $\gamma > 0$, whereas the states in the conduction edge will have $\gamma < 0$. The magnitude of $\gamma$ determines the 
size of biasing force (with $\gamma=0$ representing true AIMD forces). We have employed biased forces as an electronic constraint to model semiconductors and
insulators in our recently published work \cite{prasai2016}
where the biasing is done in just the opposite sense: to force to states out of the band gap region. 

\begin{figure}
\begin{center}
  \includegraphics[width=0.8\linewidth]{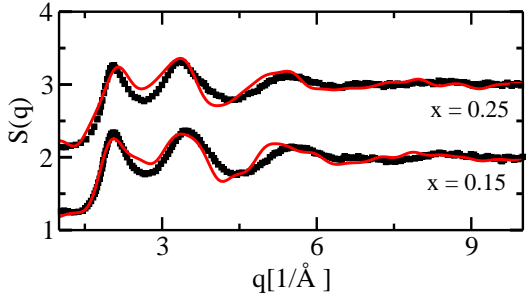}
  \caption{The structure factor of (GeSe$_3$)$_{1-x}$Ag$_x$ models (solid red line) compared with experiment
  (black squares)\cite{piarristeguy2000}}
  \label{fig1}
  \end{center}
\end{figure}

We start with conventional 240 atom models of (GeSe$_3$)$_{1-x}$Ag$_x$, $x$=0.15 and 0.25, 
at their experimental densities 5.03 and 5.31 gm/cm$^3$ \cite{piarristeguy2000} respectively. 
These were prepared using melt-quench MD simulations, followed by conjugate-gradient relaxation to a local energy minimum.
The MD simulations are performed using the Vienna {\it Ab initio} Simulation Package (VASP) \cite{kresse1,*kresse2}. 
Plane waves of up to 350 eV are used as basis and DFT exchange correlation functionals of Perdew-Burke-Ernzerhof \cite{perdew1996} were used.
Brillouin zone (BZ) is represented by $\Gamma$-point for bulk of the calculations.
For static calculations, BZ is sampled over 4 k-points.
These models fit the experimental structure factor reasonably well (Figure \ref{fig1}).

We obtain conducting conformations by annealing the starting configurations using biased forces at 700 K for 18 ps.
The electronic states in the energy range [$\epsilon_{F}$--0.4 eV, $\epsilon_{F}$+0.4 eV] are included in the computation of bias force and $\gamma = 3.0$ is used. 
The bias potential ($\Phi_{b}(R_{1},..,R_{3N})= \sum -\gamma_{i}  \langle \psi_{i}|H(R_{1},..,R_{3N})|\psi_{i} \rangle$) shepherds
the electronic states in the band edges into the band-gap region. 
Since we want any proposed metallic conformation to be a true minimum of the unbiased DFT energy functional, 
we relax instantaneous snapshots of biased dynamics (taken at the interval of 0.2 ps, leaving out the first 4 ps) to their nearest minima using conjugate gradient algorithm with true DFT-GGA forces. 
We study all relaxed snapshots by i) gauging the density and localization of states around Fermi energy and,
ii) testing the stability of the configurations by annealing them at 300 K ({\it n.b.} glass transition temperatures ($T_g$) are 488 K and 496 K for compositions $x$=0.15 and 0.25 respectively \cite{arcondo2007}).
At each composition, we selected five models that display a large density of extended states around Fermi energy and are stable against extended annealing at 300 K as the `metallized' models.
These metallized systems are, on average, 0.040$\pm$0.009 eV/atom above their insulating counterparts.

\begin{figure}
\begin{center}
  \includegraphics[width=\linewidth]{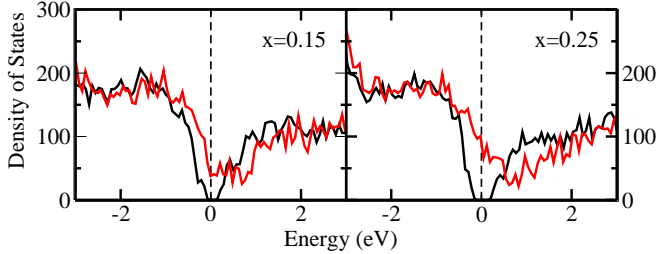}
 \caption{The electronic density of states (DOS) of the insulating model (black curve) and the metallized model (red curve).
 Energy axis is shifted to have Fermi level at 0 eV (the broken vertical line)}
  \label{fig2}
  \end{center}
\end{figure}

\begin{figure}
\begin{center}
  \includegraphics[width=\linewidth]{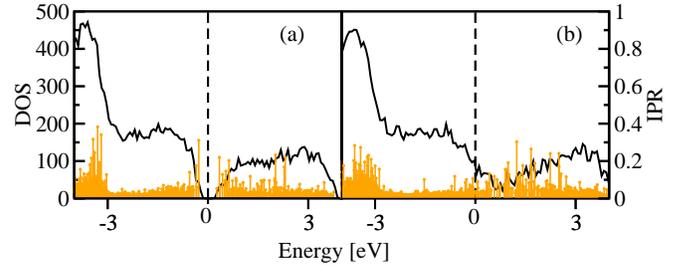}
 \caption{The (black curve) electronic density of states (DOS) and (orange drop lines) Inverse Participation Ratio (IPR) of the insulating model (a) and the metallized model (b).
 Energy axis for all datasets is shifted to have Fermi level at 0 eV (highlighted by the broken vertical line)}
  \label{fig3}
  \end{center}
\end{figure}

\begin{figure}
\begin{center}
  \includegraphics[width=\linewidth]{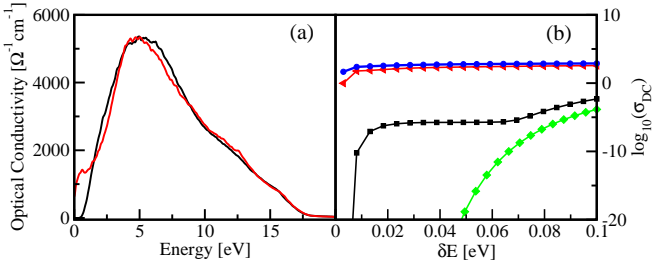}
 \caption{(a) Optical conductivity of insulating (black curve) and metallized (red curve) models for (GeSe$_3$)$_{0.75}$Ag$_{0.25}$ model computed using Kubo-Greenwood Formula (KGF). 
  Brillouin zone sampling is done over 4 k-points. Average over 3 directions was taken to eliminate artificial anisotropy. (b) DC conductivity as a function of Gaussian approximant $\delta$E.
  black squares: insulating model at $x$=015, red triangles: metallic model at $x$=0.15, green diamonds: insulating model at $x$=0.25, blue circles: metallic model at $x$=0.25}
  \label{fig4}
  \end{center}
\end{figure}

The metallized models, by construction, show a large density of states around Fermi energy (Fig. \ref{fig2}) whereas the insulating models display small but well defined PBE gap 
of 0.41 eV and 0.54 eV for $x$=0.15 and 0.25 respectively.
For disordered materials, a high DOS at $\epsilon_F$ {\it alone} may not produce conducting behaviour since these states can be localized (example: amorphous graphene, \cite{pablo}).
We gauge the localization of these states by computing inverse participation ratio (IPR, \cite{ziman})(plotted for $x$=0.25 system in Figure \ref{fig3}) and show that these states {\it are} indeed extended.
We compute the electronic conductivity [$\sigma(\omega)$] using Kubo-Greenwood formula (KGF) in the following form:
\begin{equation}\label{eq_KGF}
\begin{aligned}
{\sigma}_{k}(\omega) =  \frac{ 2 \pi e^{2} \hslash^{2}}{3 m^{2} \omega \Omega} \sum \limits_{j=1}^{N} \sum \limits_{i=1}^{N} \sum \limits_{\alpha=1}^{3}[F(\epsilon_{i},k)-F(\epsilon_{j},k)] \\
	 |\langle \psi_{j,k}|\bigtriangledown_{\alpha} |\psi_{i,k} \rangle|^{2} \delta(\epsilon_{j,k}-\epsilon_{i,k}-\hslash \omega)
\end{aligned}
\end{equation}
It has been used with reasonable success to predict conductivity \cite{abtew2007,*galli1990,*allen1987}.
Our calculations used 4 k-points to sample the Brillouin zone. To compensate for the sparseness in the DOS due to the size of the supercell, a Gaussian broadening (\dE) for the $\delta$-function is used.
We note that the choice of \dE~between 0.01 eV and 0.1 eV does not significantly alter the computed values of DC conductivity [$\sigma(\omega=0)$] (Figure \ref{fig4}). 
For the choice of \dE=0.05 eV (which is small compared to the thermal fluctuation of Kohn-Sham states for disordered systems at room temperature. For a heuristic theory, see \cite{prasai2016eph}),
the DC conductivity of metallic models are of the order of $10^{2}~\Omega^{-1}cm^{-1}$ at both concentrations.
For the insulating model at $x$=0.15, this value is of order $10^{-6}~\Omega^{-1}cm^{-1}$ whereas for insulating model at $x$=0.25, this value is lower but can not be ascertained from our calculations.
We find that the metallized models show, at least, $\sim 10^8$ times higher conductivity than the insulating models.
The computed conductivity for metallic models are comparable to the DC conductivity values of liquid silicon ($\approx 10^4~\Omega^{-1}cm^{-1}$, \cite{glazov}).

We track the atomic rearrangements associated with the metallization of network to identify the microscopic origin of metallicity.
Recalling that these are inhomogenous glasses with phase separation into Ag-rich {\it a-}Ag$_2$Se phase and residual Ge-Se backbone,
we note that the insulator-metal transition in these glasses can be viewed in terms of relative ratio of these two competing phases.
In particular, we make the following three observations associated with the insulator-metal transition:
i) Growth of Ag-Se phase, ii) Depletion of tetrahedral GeSe$_2$ phase, and iii) Growth of Ge-rich phase in host network.
Below we briefly comment on these three observations, a more detailed account of structural rearrangements will be published later.

\begin{figure}
\begin{center}
  \includegraphics[width=\linewidth]{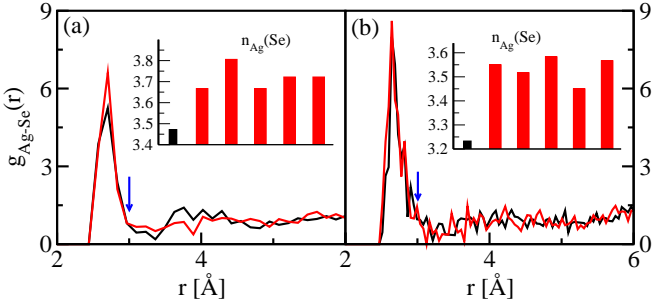}
  \caption{The Ag-Se correlation in insulating (black) and metallized (red) models at two concentrations of silver (a) $x$=0.15 and (b) $x$=0.25.
  The histogram in inset shows the Se-coordination around Ag atoms (n$_{Ag}$(Se)) for insulating (black) and 5 metallic (red) confirmations at both values of x. The cutoff
for computing coordination is 3.00 \AA, highlighted by an arrow.}
  \label{fig5}
  \end{center}
\end{figure}

\begin{figure}
\begin{center}
   \includegraphics[width=.8\linewidth]{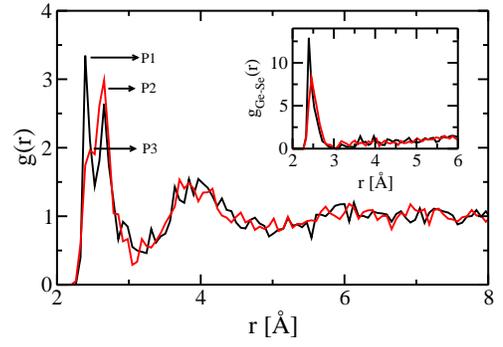}
  \caption{The total radial distribution function (g(r)) of the insulating and metallized models (black and red curves respectively) at $x$=0.25. 
 Note the bifurcated first peak originates from Ge-Se correlation (P1 at 2.40 \AA) and Ag-Se correlation (P2 at 2.67 \AA). 
 For metallized model, peak P3 arises due to depletion of tetrahedral \gese2 and formation of Ge-rich Ge-Se phases.} 
  \label{fig6}
  \end{center}
\end{figure}

{\it Growth of Ag-Se phase.} We observe that the Ag-Se phase grows upon metallization. 
Se-Ag correlation ($r_{Ag-Se}=2.67$ \AA) is found to increase from the insulating to metallic model (see Figure \ref{fig5}, also the increase in peak P2 in Figure \ref{fig6}).
For both Ag concentrations, Se-coordination around Ag is found to increase from insulating to metallic models.
For $x$=0.15, Se-coordination around Ag increases from 3.47 to 3.72 (the later value is an average over 5 metallic models, see Figure \ref{fig5}). 
For $x$=0.25, it increases from 3.23 to (on average) 3.53.

{\it Depletion of tetrahedral GeSe$_2$.} 
The network in the insulating phase is dominated by Se-rich tetrahedral \gese2, accompanied by a competing Ag-Se phase.
The fraction of later phase is directly determined by Ag-concentration in the network.
These two phases appear as two distinct peaks in total radial distribution function (RDF) (Figure \ref{fig6}).
Upon metallization, the growth of Ag-Se shifts the balance of stoichiometries in network and the host network becomes Se deficient.
At composition $x$=0.25 (plotted in Fig. \ref{fig6}), the network in metallic phase is dominated by the Ag-Se subnetwork (peak P2).
The corresponding Ge-Se coordination number in metallic model is 3.22, slightly lower than 3.40 in insulating model.
These values are 3.28 and 3.43 respectively for $x$=0.15. The correlation cutoff of 2.70 \AA~is taken to determine the coordinations.

{\it Response of host network}. 
The host network of Ge-Se consists of Se-rich tetrahedral GeSe$_2$ and non-tetrahedral Ge-rich phases including the ethane-like Ge$_2$Se$_3$ units.
These subnetworks were reported in GeSe$_2$ by Boolchand and coworkers \cite{boolchand2000} and in ternary chalcogenide glasses by Mitkova and coworkers \cite{mitkova2006}.
We find that these Ge-Se stoichiometries have different bondlength distributions: Se rich phases ($n_{Ge-Se} \geq 4$) have bondlengths smaller than 2.55 \AA~whereas 
Ge-rich phases ($n_{Ge-Se} < 4$) have bondlengths longer than 2.55 \AA.
In an insulating conformation, the former phase dominates and registers an RDF peak at $\approx$ 2.40 \AA~(Fig. \ref{fig6}). 
For metallic conformations, fewer Se atoms are available to Ge.
This increases the fraction of Ge-rich phases and the Ge-Se bondlength distribution shifts to longer distances.
This is represented by a shift in Ge-Se pair correlation function in Figure \ref{fig6} (inset) and appearence of peak P3 in total RDF.  
Due to increase in fraction of Ge$_2$(Se$_{1/2}$)$_6$, Ge-Ge correlation peak appears around 3.5 \AA~in metallic models. 
We note that it is such a Ge-Ge signal in Raman scattering and $^{119}$Sn M$\ddot{o}$ssbauer spectroscopy that led to experimental
discovery of Ge-rich Ge$_2$(Se$_{1/2}$)$_6$ phase in stoichiometric bulk Ge$_x$Se$_{100-x}$ glasses \cite{boolchand2000}.

\begin{figure}
\begin{center}
  \includegraphics[width=0.8\linewidth]{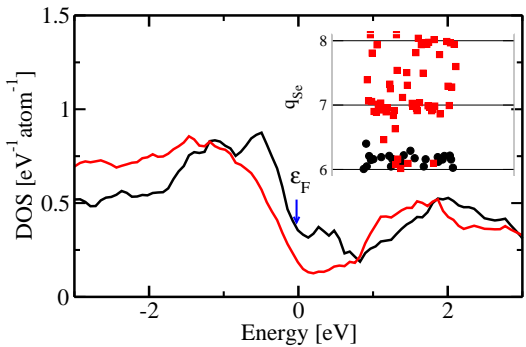}
  \caption{The density of states of metallic model projected onto Se-atoms in the two subnetworks: Ag-Se subnetwork (black curve) and Ge-Se subnetwork (red curve).
  Since these two subnetworks contain different number of Se atoms (23 and 59 for this plot), an average was taken to enable comparision.
  Bridging Se-atoms are not included in the calculation. Energy axis was shifted to have Fermi energy ($\epsilon_{F}$) at 0 eV. The inset shows 
  Bader charges (q$_{Se}$) for the same two groups of Se atoms. Black filled circles represent Se in Ag-Se network, Red filled squares represent Se in Ge-Se network.}
  \label{fig7}
  \end{center}
\end{figure}

Now we comment further on the role of Ag-Se phase in metallicity.
It is well known that the states around Fermi energy are mainly Se p-orbitals (\cite{prasai2011,tafen2005}, In GeSAg: \cite{akola2015}).
The electronic structure of metallic model projected onto its constituent subnetworks (Ag-Se and Ge-Se) shows different electronic activity of Se-atoms in the two subnetworks.
We find that {\it individual} Se atoms in Ag$_2$Se nework have twice as much projection around the Fermi energy than the Se atoms in Ge-Se network (Figure \ref{fig7}).
This suggests that a more concentrated Ag-Se network will enhance the conduction. 
Experimentally, growth of Ag-rich nanocrystals in GeS$_2$ matrix has been shown to enhance the {\it electronic} conductivity \cite{wang2012,*waser2009}.
The Se-atoms in Ag-Se phase are found in atomic state ($q_{Se} \sim 0$) where as those in Ge-Se network are ionic with 
negative charge ($q_{Se} \sim -1$ or $-2$) (See inset in Figure \ref{fig7}).

Altogether, we have presented a direct simulation of conducting phase of CBRAM material GeSe$_3$Ag and
it provides evidence of the conduction through interconnected regions of Ag$_2$Se phase in the glassy matrix \cite{wang2012, *waser2009}.
This work does not attempt to describe the conduction through Ag-nanowires which may be the mechanism of conduction in two terminal metal-electrolyte-metal devices \cite{inbook1}.
It demonstrates the existence of metastable amorphous forms (``poly-amorphism") of the glass with drastically different optical response.
The observation that the DFT energies of these states are only 0.04 eV/atom higher than insulating state suggests that these states might to accessible.
Furthermore, we have shown that ``Gap Sculpting" can be used to purposefully {\it design} metallic conformation.

We thank M. Mitkova and P. Boolchand for stimulating discussions.
This work is supported by National Science Foundation under grant no. DMR 1506836, no. DMR 1507670 and no. DMR 1507166.
We are thankful to Ohio Supercomputer center for computing resources.

\end{document}